\documentclass{emulateapj}
\usepackage{apjfonts}
\usepackage{natbib}
\usepackage{graphicx}
\bibpunct{(}{)}{;}{a}{}{,}% magic!
\newcommand\linea{Fe~{\sc i}~$\lambda$6301.5~\AA}
\newcommand\lineb{Fe~{\sc i}~$\lambda$6302.5~\AA}
\newcommand\linec{Fe~{\sc i}~$\lambda$15648~\AA}
\newcommand\lined{Fe~{\sc i}~$\lambda$15652~\AA}
\newcommand\linee{Sr~{\sc i}~$\lambda$4607~\AA}
\slugcomment{}%ver3}
\shorttitle{Magnetic field strength PDF of the quiet Sun}
\shortauthors{Dom\'inguez Cerde\~na, S\'anchez Almeida \& Kneer}

\begin{document}

%\date{ }
\title{
	The distribution of Quiet Sun
	magnetic field strengths from 0 to 1800~G
}

   \author{I. Dom\'\i nguez Cerde\~na}
   \affil{Instituto de Astrof\'\i sica de Canarias,
	      E-38205 La Laguna, Spain}
   \email{itahiza@iac.es}

    \author{J. S\'anchez Almeida}
    \affil{Instituto de Astrof\'\i sica de Canarias,
	      E-38205 La Laguna, Spain}
   \email{jos@iac.es}

%    \and
%    \author{F. Kneer}
%    \affil{Universit\"ats-Sternwarte,
%	      Geismarlandstra\ss e 11, D-37083 G\"ottingen, Germany}
%  \email{kneer@uni-sw.gwdg.de}
    \and
    \author{F. Kneer}
    \affil{Institut f\"ur Astrophysik,
	      Friedrich-Hund-Platz 1, D-37077 G\"ottingen, Germany}
   \email{kneer@astro.physik.uni-goettingen.de}

\begin{abstract}
The quiet Sun photospheric plasma has a variety 
of magnetic field
strengths going from zero to 1800 G.
The empirical characterization of 
these field strengths requires a probability density 
function (PDF), i.e., a function $P(B)$ describing
the fraction of quiet Sun occupied
by each field strength $B$.
We show how to combine magnetic field strength
measurements based on the Zeeman effect and
the Hanle effect to estimate an unbiased
$P(B)$.
The application of the method to
real observations
renders a set of  possible PDFs, which outline
the general characteristics 
of the quiet Sun magnetic fields.
Their most probable field strength differs
from zero.
The magnetic energy density is a 
significant fraction of the kinetic energy
of the granular motions at the base of the
photosphere (larger than 15\% or larger than 
$2\times 10^{3}~{\rm erg~cm}^{-3}$).
The unsigned flux density 
(or mean magnetic field strength) 
has to be between 130~G and 190~G.
A significant part of the unsigned
flux (between 10\,\% and 50\,\%)
and of the magnetic energy  
(between 45\,\% and 85\,\%) are provided
by the field strengths larger than
500~G which, however, occupy only a small fraction
of the surface (between 1\% and 10\%).
The fraction of kG fields in the quiet Sun 
is even smaller, but they are important
for a number of reasons.
The kG fields still trace a  significant fraction 
of the total magnetic
energy, they reach the high  photosphere, 
and
they appear in unpolarized light images. 
The quiet Sun photosphere has far more unsigned magnetic 
flux and magnetic energy than
the active regions and the network
all together.
\end{abstract}

\section{Introduction}\label{intro}

The interior of the supergranulation cells does not 
seem to be magnetic according to the standard full
disk magnetograms. 
\citet{liv75} and \citet{smi75} improved the 
standard sensitivity to discovered  weak
magnetic signals in the cell interiors. These
elusive signals are known as Inter-Network 
fields (IN),  Intra-Network
fields or, simply, quiet Sun fields. 
Characterizing these quiet Sun fields is  
important since they may play a  role in the
physics responsible for the global solar
magnetism \citep[][]{unn59,
ste82,yi93,san98c,san04,sch03b}.
The quiet Sun could easily contain
most of the photospheric 
unsigned magnetic flux and magnetic energy,
simply because it  occupies most of the solar surface
(more than 90~\%, even at solar maximum; 
\citealt{har93}).

Our understanding of the quiet Sun magnetic fields
has improved during the last years thanks to 
advances in instrumentation
\citep[e.g.][]{gro96,san96,sig99,lin99,lit02,
dom03a,dom03b,kho02,san04a}, 
diagnostic capabilities 
\citep[e.g.][]{ste82,fau93,fau95,lan98,san00,soc02,
tru04,man04,san05}, 
as well as  numerical simulations 
\citep[e.g.][]{cat99a,emo01,ste02,vog03b,vog05}.
The new spectropolatimeters 
make it possible to detect the low polarimetric signals
characteristic of the IN fields.
Sophisticated diagnostic techniques are needed to
interpret those signals, which depend in a non-trivial
way on the (a priori unknown) physical properties of the 
magnetized plasma. For example,  opposite 
polarities often coexist in the resolution elements giving 
rise to very asymmetric line profiles requiring 
special methods of diagnostic
\citep[see, e.g.,][]{san00,lit02}.  
The numerical simulations provide guidance for the
magnetic field topology to be expected, a
knowledge critical  for a proper 
interpretation of the observables.

The magnetic field strength is one of the
key parameters  to characterize the magnetism
of the quiet Sun. Its study has a rather long tradition,
with a period of unsuccessful attempts 
\citep[][]{unn59,how69,ste77},
followed by the first estimates by \citet{ste82}
and \citet{fau93} based on Hanle depolarization 
measurements.
The observed Hanle signals agree with a
volume-filling mixed-polarity magnetic field of 
constant field strength 
\citep{ste82,fau93,fau95,fau01,ste97b}. The value
of the estimated mean field strength has increased 
upon refinement of the diagnostic techniques.
The most recent estimate by \citet{tru04} and
\citet{bom05} are
consistent with a constant field of 60~G (or a mean field of
130~G assuming an exponential distribution
of magnetic field strengths). Most measurements 
of the quiet Sun magnetic fields
are based on interpreting line polarization signals
caused by the more familiar Zeeman effect. They show
a rather different magnetism, with structures occupying a
small fraction of the solar surface and having
intense  fields in the range of the hectogauss (hG) and 
the kilogauss (kG). Actually, the field strength
depends in a systematic way on the Zeeman splitting
of the spectral line employed to measure. The 
works using visible lines deduce the
presence of kG magnetic fields
\citep[][]{gro96,san00,soc02,dom03a,dom03b}, 
while measurements based on infrared (IR)
lines infer sub-kG field strengths \citep[][]{lin99,kho02}.
This apparent inconsistency among the various measurements
of field strength, plus the guidance of numerical simulations
of magneto-convection, suggest that the quiet Sun 
has a continuous distribution of field strengths
going all the way from 0~G to 2~kG
\citep{san00,soc03,tru04}. Therefore,  
each one of the individual measurements
mentioned above  is biased, providing only a
partial view of the
magnetic fields existing in the quiet Sun. We think
that  this biased information can be cleaned up
and assembled to provide a reasonable
estimate of the intrinsic magnetic field strength
distribution. 
This work describes a method to recover such 
information.

Due to the 
variety of field strengths,  they are best characterized
by a probability density function PDF.  The magnetic field
strength PDF, $P(B)$, is defined as the probability 
that a point of the atmosphere chosen at random
has a magnetic field strength $B$. Alternatively, it describes
the fraction of quiet Sun occupied by fields with a 
strength $B$. The PDF approach to characterizing
the quiet Sun field strength is both convenient and
powerful.
First,
one can directly compare observations 
with numerical simulations,
which predict $P(B)$. Second,
it concentrates basic information on the quiet Sun magnetism,
e.g., the first moment 
$\langle B\rangle$ is 
connected with the unsigned magnetic flux 
density\footnote{
Strictly speaking, the unsigned magnetic flux 
density across a  plane is given by the
average of the unsigned component of the magnetic
field in the direction perpendicular to the plane.  
Consequently, we are not precise when using the term
{\em unsigned magnetic flux 
density} for $\langle B\rangle$, which
is the average of the magnetic field 
vector modulus.  
There is  no possible misunderstanding, though.
The term is employed consistently throughout the text,
and the actual unsigned flux is never mentioned
except in this footnote.
}, 
whereas the
second moment provides the magnetic energy 
density
$\langle B^2\rangle/(8\pi)$,
\begin{equation}
\langle B\rangle=\int_0^{\infty}B\, P(B)\,dB,
\label{mean}
\end{equation}
\begin{equation}
{\langle B^2\rangle}=\int_0^{\infty}B^2\, P(B)\,dB.
\label{mo1}
\end{equation}
There are several observational magnetic field 
strengths PDFs in the literature (see the references
cited in the previous paragraph). As we pointed out above, all of
them are known to be severely biased. 
On the one hand,
those based on Hanle
signals miss the  strong fields,
since the Hanle depolarization does not 
vary for field strengths above a threshold
that is typically smaller than a few hundred~G 
\citep[see, e.g.,][]{lan92,ste94,tru01}.
On the other hand, the Zeeman measurements show only
a small fraction of the existing magnetic structures
\citep[see, e.g.,][]{ste82,san03}.
The Zeeman signals tend to cancel when opposite polarities
coexist in the resolution elements. We detect only 
the residuals left when the cancellation is not perfect
(e.g., when the two polarities do not have exactly the
same magnetic flux, when strong 
Doppler shifts separate the absorption produced by the
two polarities, when the
temperature imbalance between polarities 
creates net signals, and so on).
Here  we try to
get rid of all these observational biases
to produce an empirical $P(B)$. We approach the
problem by assuming a flexible yet reasonable shape
for the expected PDF. This guess is based on 
recent numerical simulations of
magneto-convection, as well as on the
latest Hanle and Zeeman observations. The 
actual $P(B)$ is  set by forcing the PDF to
meet several observational and theoretical
constraints. So far as we are aware of, this is
the first quiet Sun PDF satisfying
magnetic field strength 
observations going all
the way from 0~G to 2~kG. (The 
premises were already put forward by \citealt{san04},
though.)
The success of the trial is only partial since rather than
a single  PDF we derive a family compatible with the
present
observations. However, the approach is shown to be
viable, so that future more precise observations
will allow us to narrow down the range of
possibilities. 

The paper is organized as follows;
\S~\ref{synthesis} describes how to piece together all the
ingredients needed to estimate an unbiased PDF.
Some of the more tedious details are completed in 
Appendix~\ref{ap1}.
The properties of the PDFs satisfying the
observational constraints are extracted and 
analyzed in \S~\ref{results}.
In particular, we compare the semi-empirical PDFs with
theoretical PDFs coming from numerical simulations of 
magneto-convection.
Finally, \S~\ref{conclusions} summarizes the
main conclusions, discusses their consequences,
and points out the natural way to proceed.

%%%%%%%%%%%%%%%%%%%%%%%%%%%%%%%%%%%%%%%%%%%%%%%%%%%%%%%%%%%%%%%%%%%%

\section{Synthesis of the PDF}\label{synthesis}
 
In order to facilitate the reading
of this detailed section, we start off by listing
the thread of the argument.

\begin{enumerate}
\item The {\em unbiased} quiet Sun PDF can be approximated
by a linear combination of the {\em biased} PDFs inferred from
Zeeman  signals and from Hanle signals (\S~\ref{shape}).
\item The shape of the Zeeman PDF is obtained from observations.
The shape of the Hanle PDF is not constrained by observations, and
so we  take it from numerical simulations of 
magneto-convection (\S~\ref{shape}).
\item The Hanle and Zeeman signals come from different photospheric
heights. One  needs a transformation to express them in a 
common height, in other words, a  recipe for the variation 
with height in the atmosphere of the PDFs (\S~\ref{strat}).
\item Piecing together this information, we work out a 
semi-analytical PDF depending on three free parameters.
One needs only three independent observables to set
it (\S~\ref{constraints}). 
\end{enumerate}

\subsection{The shape of the PDF}\label{shape}
The Zeeman induced polarization changes the sign depending on the
orientation of the vector
magnetic field with respect to the observer.
When magnetic fields with varied directions coexist in the
resolution elements, the net polarization tends to cancel. The
lack of polarization is interpreted as  absence of 
magnetic fields, so that the Zeeman  measurements 
underestimate the volume occupied by magnetic fields.
This bias increases with decreasing  field strength, since
the Zeeman signals scale with the field strength. 
The Hanle induced depolarization
does not suffer the cancellation of the Zeeman signals.
However, the Hanle induced depolarization is
insensitive to magnetic field strengths larger than
a few hundred~G
\citep[e.g.,][]
{lan92,ste94,tru01,san05}. 
Thus the biases of the Hanle and Zeeman measurements
turn out to be complementary.   
Keeping in mind this complementarity, one can approximate the true
unbiased PDF, $P(B)$, as a linear combination of the
biased PDF inferred from Zeeman signals, $P_Z(B)$, plus
the biased PDF that fits the Hanle signals, $P_H(B)$,  
\begin{equation}
P(B)=w\,P_H(B)+(1-w)\,P_Z(B),
\label{eq1}
\end{equation}
with $P_H(B)$ and $P_Z(B)$ normalized to unity, and 
the weight $w\simeq~1$. 
A full derivation of this equation is given in Appendix~\ref{ap1}.

\begin{figure}
\resizebox{\hsize}{!}{\includegraphics{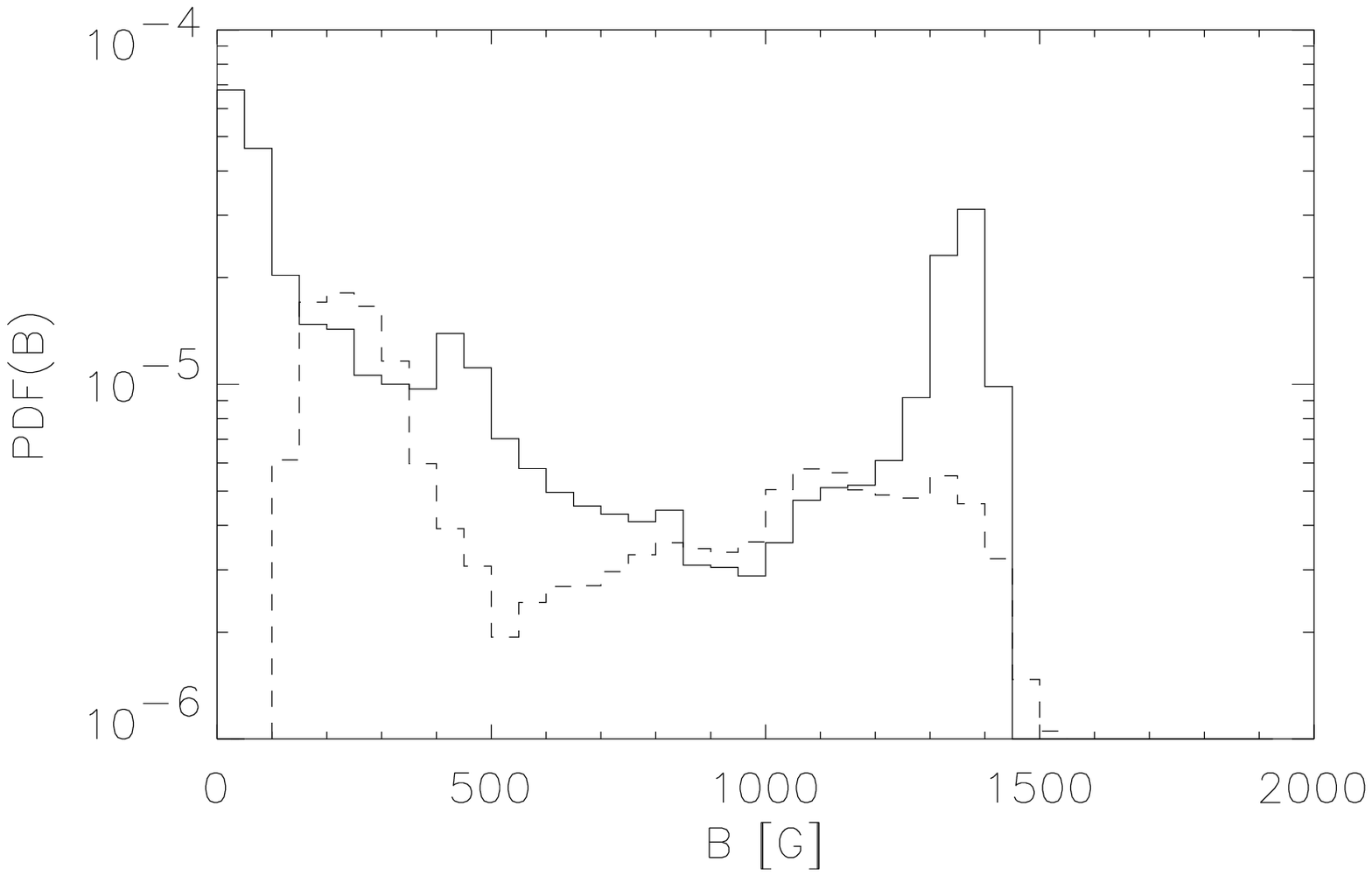}}%pdf0.ps}}
\caption{PDFs inferred from Zeeman signals
of \linea, \lineb, \linec, and \lined\
observed simultaneously in a quiet Sun region.
The two PDFs  come from the same data
set. The solid line represents a MISMA
inversion \citep{dom05} whereas the
dashed line corresponds to a Milne-Eddington
inversion \citep{san03c}. The MISMA
PDF is shown some 100~km above the base of the
photosphere, at a height corresponding
to the Milne-Eddington inversion. The PDFs
are normalized to the fraction
of quiet Sun producing 
Zeeman polarization signals ($\sim$\,1.5\,\%).
}
\label{pdfz}
\end{figure}

Regarding $P_Z(B)$, we take the shape of the biased PDF measured
by \citet{dom05}. 
It is based on a realistic semi-empirical modeling of 
2280
Stokes profiles of \linea, \lineb, \linec, and \lined\
observed simultaneously in a quiet Sun region. These
four spectral lines are complementary 
since those in the
visible spectral range sense preferentially 
the kG fields whereas the IR lines tend to detect hG
field strengths \citep{san00,soc02,soc03}.
The empirical Zeeman PDF shape adopted here
is not very different from 
the one derived by means of
Milne-Eddington fits to the same Stokes profiles 
\citep[see][]{san03c,dom04}.
The two PDFs are shown in Figure~\ref{pdfz}.
Both are rather broad with a local maximum
at  kG fields. The differences are 
unimportant for the present work, and they
are discussed in detail by \citet{dom05}.

\begin{figure}
\resizebox{\hsize}{!}{\includegraphics{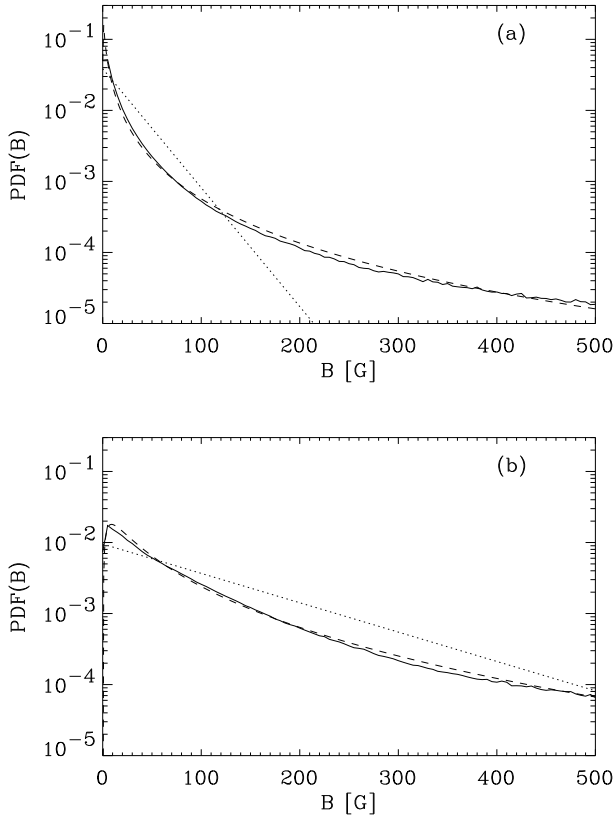}}%pdf1.ps}}
\caption{Fits of quiet Sun PDFs from 
	numerical simulations of magneto-convection \citep[the solid
	lines;][]{vog03b} using log-normal functions (the dashed
	lines) and exponential functions (the dotted lines).
	The log-normals are far superior. The upper and lower panels
	differ in the initial field  strength
	of the simulation; 10~G (a) and 50~G (b),
	respectively. The PDFs correspond to the base of the
	photosphere.
	}
\label{pdf1}
\end{figure}

The Hanle effect modifies the polarization produced
in scattering processes, and so it only affects lines
formed in  NLTE (non-local thermodynamic equilibrium). 
Hanle based studies are therefore restricted
to lines formed high in the atmosphere,
where the collisions  are low enough for the scattering 
events to be frequent.
However, we need Hanle signals formed as low as possible,
so that they can be compared with Zeeman signals
produced in the low-mid photosphere.
The Hanle signals of \linee\ are formed particularly low
\citep[e.g.][]{fau95}. In addition, this line allows a simple
treatment of the usually very complex synthesis of the 
Hanle signals  \citep[see, ][and also the discussion in 
\S~\ref{constraints}]{fau01}. 
\linee\ is chosen to constrain
the PDF for these two unusual properties. As it is argued in, e.g.,
\citet{san05}, the Hanle signals of \linee\ do not
constrain the  shape of the PDF.  This limitation is
overcome here by parameterizing the shape of $P_H(B)$ using an 
analytic function. The Hanle signals have 
been interpreted with approximations
using Dirac $\delta$-functions, exponential functions,
and Maxwellian distributions
\citep[e.g.,][]{ste82,fau93,tru04,san04}, but these shapes
do not fit in our needs. First, 
we expect a continuous distribution of field strengths,
which discards Dirac $\delta$-functions. Second,
the exponential fall off with field strength,
\begin{equation}
P_H(B)=\frac{1}{\langle{B}\rangle}\exp\big({-B/\langle{B}\rangle}\big),
\label{exp}
\end{equation}
is too pronounced according to the PDFs resulting from
numerical simulations of magneto-convection 
(see, e.g., the dotted line in Figure~\ref{pdf1}a).
The Maxwellian distribution can be discarded with the
same argument, since the fall off is even more
pronounced than the exponential one.
Inspired by the numerical
simulations of magneto-convection
developed with the MuRAM code 
\citep{vog03b,vog05},
we choose a log-normal distribution $P_H(B)$,
\begin{equation}
%P_H(B)=\frac{1}{\sqrt{\pi}\sigma B}e^{-(Ln(B)-Ln(B_0))^2/\sigma^2},
P_H(B)=\frac{1}{\sqrt{\pi}\sigma B}
\exp\big[-(\ln B-\ln B_0)^2/\sigma^2\big],
\label{logn}
\end{equation}
with $B_0$ and $\sigma$ two  parameters 
related to the
two first moments of the distribution, 
\begin{equation}
\langle B\rangle=B_0\exp(\sigma^2/4),
	\label{firstmoment}
\end{equation} 
and
\begin{equation}
\langle B^2\rangle=B_0^2\exp(\sigma^2).
	\label{secondmoment}
\end{equation}
%
%Fig 1%%%-------------------------
%
Figure~\ref{pdf1} 
compares  numerical PDFs from \citet{vog03b} with fits
using  log-normal functions and exponential
functions. (All the fits were 
performed for $B <$ 500~G.) The two plots correspond to an initial
vertical field of 10~G (Figure~\ref{pdf1}a) and  
50~G (Figure~\ref{pdf1}b),
and the parameters of the log-normal PDFs
are $\sigma=2.5,\,B_0=5~$G, and $\sigma=1.7,\, B_0=38~$G, respectively.
The log-normal fits are much better than the
exponentials, and we choose them to represent $P_H(B)$.
A comment is in order. The log-normals tend to zero when
$B\rightarrow$0~G. Actually, they reach a maximum at $B=B_{\rm max}$
with,
\begin{equation}
B_{\rm max}=B_0\,\exp(-\sigma^2/2).
\label{maxposition}
\end{equation}
Counterintuitive as it sounds, 
the property of having a maximum is to be expected for the 
true quiet Sun
PDF. When the field strength is small enough, the magnetic
fields are easily modified and randomized 
by the convective motions. Those weak fields
are expected to have an isotropic distribution of field 
directions, with the three components of the
magnetic field being independent. Suppose that the 
magnetic field strength of these weak
fields has a typical value $B_t$ different
from zero but
otherwise arbitrary. For a point of the atmosphere to 
have $B=0$ the three components of the magnetic field
vector must be zero simultaneously. This event is highly 
improbable since the three components are independent.
Consequently, it is to be expected that $P(B)\rightarrow$0
when $B << B_t$. 
The drop at small fields is also clearly seen in some numerical 
simulations of magneto-convection 
(e.g., Figure~\ref{pdf1}b; Figure~3 in \citealt{cat99a}; 
Figure~5.25 in \citealt{vog03b}).

%%%%%%%%%%%%%%%%%%%%%%%%%%%%%%%%%%
\subsection{Vertical variation of the PDF}\label{strat}

The photosphere is strongly stratified, with the gas
pressure decreasing exponentially with the height in the
atmosphere. Part of the pressure that  maintains the
mechanical balance of a magnetic structure is 
provided by the magnetic pressure, which has to
decrease to keep up with the pressure drop.
Since the magnetic flux is conserved, the decrease
of field strength comes together with an expansion
of the field lines and an increase of the volume
occupied by the magnetic fields 
\citep[see, e.g.,][]{spr81}.
These two effects modify the PDF. 
As we pointed out above,
the Hanle signals are formed high in the photosphere.
In order to use them to constrain a PDF at the base
of the photosphere, one needs to know how 
the PDF varies with height.
Working out such variation 
is a non trivial problem, closely connected
with the so-called extrapolation of the photospheric
fields to the chromosphere and corona 
\citep[e.g.,][]{sak89,sch03b}.
Here we develop a crude  approximation, which suffices
for the kind of exploratory investigation carried out in the
paper but which, eventually, will require further refinement. 

\begin{figure}
\resizebox{\hsize}{!}{\includegraphics[angle=-90]{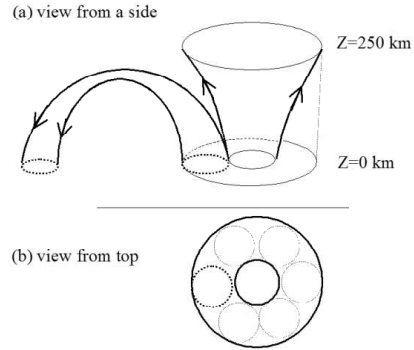}}%height_var.ps}}
\caption{(a) Cartoon with two magnetic fluxtubes. The 
field lines of one of them reach the mid photosphere
($z=250$~km). The field lines of the second one
bend over and return below this photospheric height. 
(b) A view from the top of the first fluxtube.
The thick solid  circles outline  the sections
of the fluxtube at $z=0$~km (small circle) and $z=250$~km
(large circle). Note that for the fluxtube to expand
a number of other concentrations around it  
cannot reach $z=250$~km. The section
of those that must return are represented by dotted 
 circles in the cartoon. 
}
\label{height_var}
\end{figure}

We start off by introducing a new symbol for the PDF at a height
in the atmosphere $z$, $\wp(B,z)$, so that
the PDF discussed so far corresponds to 
the base of the photosphere\footnote{
The height $z=0$ corresponds to  the 
base of the photosphere in the one-dimensional
mean quiet Sun model atmospheres \citep[e.g.][]{mal86},
where the continuum 
optical depth $\tau_c=1$and the total pressure is 
$\simeq 1.2\times 10^{5}$~dyn~cm$^{-2}$.},
\begin{equation}
P(B)=\wp(B,0).
\label{defpz}
\end{equation}
Our estimate is based in the so-called thin fluxtube approximation
\citep[e.g.,][]{spr81,spr81b}. 
In this approximation, the magnetic field
strength of an isolated static structure surrounded by field free
plasma drops with the height in the 
atmosphere as the square root of the field free
gas pressure. In 
our case,
\begin{equation}
B(z)\simeq B(0)\,\exp(-z/H_B),
\label{b1}
\end{equation}
with magnetic field strength scale height $H_B$ twice the gas
pressure scale height, and so, independent of the field strength. 
The cross-section of one of these thin fluxtubes
$S$
increases with height as the field strength decreases. The need to
conserve the magnetic flux imposes,
\begin{equation}
S\big(B(z),z\big)=S\big(B(0),0\big)\,\exp(z/H_B),
\label{b2}
\end{equation}
so that the product $B\,S$ does not depend on $z$.
The expansion is made at
the expense of the field free plasma surrounding the
magnetic concentration. 
In our case there is no such 
field free plasma, so that an increase of the volume
occupied by one magnetic structure requires
shrinking the volume occupied by others 
(see the two magnetic
concentrations in Figure~\ref{height_var}a).
Which physical parameter determines why some 
concentrations expand whereas others shrink? 
The magnetic forces responsible for the expansion
or contraction scale with the square of the magnetic
field strength. Then it is to be expected that the
expansion of the strongest fields prevails, making it
difficult for the weak fields to reach high photospheric
layers.   
We use this idea as an ansatz to proceed.
For a magnetic concentration
to expand from $S(B(0),0)$ to $S(B(z),z)$,
its field strength has  to be the largest in the
area $S(B(z),z))$ around it (see Figure~\ref{height_var}b).
If in average all magnetic concentrations have the same 
section, then a concentration
will survive if and only if it has the largest field
strength among the $n-1$ concentrations around it, with
\begin{equation}
n\simeq{{S\big(B(z),z\big)}\over{S\big(B(0),0\big)}};
	\label{nn}
\end{equation}
see Figure~\ref{height_var}b.
The probability that one  concentration
has a field strength smaller than $B$ at $z=0$ is,
\begin{equation}
M(B)=\int_0^B P(B')\,dB'.
\end{equation}
Since the field strengths of different concentrations are
independent, the probability that $n-1$  have fields
smaller than $B$ is $M(B)^{n-1}$ and, consequently, a 
magnetic field  strength $B$ reaches $z$ with a probability,
\begin{equation}
F_P(B)=M(B)^{n-1}.
\label{FP}
\end{equation}
This probability of survival $F_P(B)$ allows us
to write down the PDF at $z$, namely,
\begin{equation}
\wp\big(B(z),z\big)dB(z)=
{{S\big(B(z),z\big)}\over{S\big(B(0),0\big)}}
F_P\big(B(0)\big)
\wp\big(B(0),0\big)\,dB(0),
\label{var1}
\end{equation} 
where we have taken into account that the atmosphere occupied
for those magnetic fields that survive increases as 
$S\big(B(z),z\big)/S\big(B(0),0\big)$.
Using equations~(\ref{defpz}), (\ref{b1}), (\ref{b2}),
and (\ref{nn}), one can rewrite equation~(\ref{var1}) in
a compact way,
\begin{equation}
\wp\big(B,z\big)=
n^2\,
P(nB)\, \big[\int_0^{nB}P(B')\,dB'\big]^{n-1}.
\label{var2}
\end{equation} 
It is not difficult to show that $\wp(B,z)$ is properly 
normalized even for non-integer $n$, which we interpret
as a sign of self consistency in the derivation 
of equation~(\ref{var2}).

Equation~(\ref{var2}) provides the PDF at any height given $n$
and the PDF at height  equals zero. We apply such recipe to
the Hanle signals of \linee\ when they are formed as close
as possible to the disk center (i.e., as low as possible 
in the photosphere). This lowest height of formation 
turns out to be about 250~km \citep{fau95}. On the other hand, 
a $H_B\simeq 230~$km is 
obtained from the stratification of $B$ in the inversions
performed by \citet{dom05}, 
which renders,
\begin{equation}
n=\exp(250~{\rm km}/H_B)\simeq 3.
	\label{neq3}
\end{equation}

%%%Fig2%%-------------------------------

\begin{figure}
\resizebox{\hsize}{\hsize}{\includegraphics{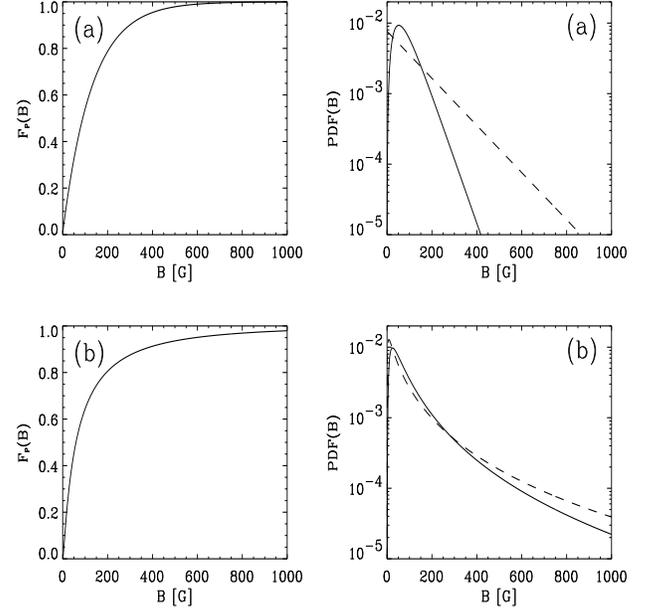}}%pdf2.ps}}
\caption{
	Examples of the change with height in the photosphere
	of an exponential PDF (a) 
	and a log-normal PDF (b). The plots in the first column show the 
	probability of survival $F_P(B)$. The plots in the
	second column represent the PDF at $z=0$ 
	(the dashed lines),
       and the PDF at z=250~km (the solid lines). (The PDFs correspond to those shown 
	in Figure~\ref{pdf1}b.)
  }
\label{pdf2}
\end{figure}

Figure~\ref{pdf2} shows 
the variation with height predicted by 
equation~(\ref{var2}) for
an exponential $P(B)$ (a), and a log-normal $P(B)$ (b). One can compare
the original PDFs (the dashed lines) with
the PDFs at $z=250$~km (the solid lines).
Note how the maximum of $P(B)$ is shifted to
stronger fields as one 
moves up in the photosphere. This is to be expected 
since the probability of survival of weak 
fields is very low.

\subsection{Physical constraints}\label{constraints}

Once $P_Z(B)$ has been set according to the prescription
in \S~\ref{shape},
The analytical PDF in equation~(\ref{eq1}) depends on three free
parameters; the weight $w$, and the two parameters $\sigma$ and
$B_0$ defining the shape of the Hanle PDF 
(equation~[\ref{logn}]). 
They can be set using three {\em independent} observables.
Unfortunately, only one observable plus one
lower limit are available. First, the fraction of the
photosphere producing the observed Zeeman signals is a few 
per cent \citep[e.g.][]{lin99,san00,dom03a,lit04b}.
This fraction represents a lower limit to the true
percentage of quiet Sun with strong fields, which
constrains $1-w$  (equation~[\ref{comment_eq}]).
In particular,
the signals used to estimate $P_Z(B)$ arise from 
some 1.5~\% of the surface \citep{dom05}. Taking 
this value as our lower limit,
$1-w > 1.5\times 10^{-2}$, and so, 
\begin{equation}
w < 0.985.
\label{ww}
\end{equation}
Second, the Hanle depolarization signals of \linee ,
$Q/Q_0$, depend on the PDF approximately as,
\begin{equation}
Q/Q_0= \int_0^{\infty}\wp(B,z)\,W_B(B)\,dB,
\label{Hcond2}
\end{equation} 
with the kernel $W_B(B)$ given by,
\begin{equation}
W_B(B)=1-{{2}\over{5}}\Big({{\gamma_H^2}\over{1+\gamma_H^2}}
+{{4\gamma_H^2}\over{1+4\gamma_H^2}}\Big),
\end{equation}
\begin{displaymath}
\gamma_H={{B}\over{50~{\rm G}}};
\end{displaymath}
see, \citet{fau01}, \citet{san05}, and references therein.
Using equations~(\ref{var2})
and (\ref{neq3}), equation~(\ref{Hcond2}) can be rewritten
as,
\begin{equation}
Q/Q_0\simeq 9\int_0^{\infty}W_B(B)\,P(3B)\, 
\big[\int_0^{3B}P(B')\,dB'\big]^2\,dB.
\label{Hcond}
\end{equation}
The work by \citet{tru04} gives our best 
estimate of $Q/Q_0$,
\begin{equation}
Q/Q_0=0.41\pm0.04,
\label{depo}
\end{equation}
with the error bars representing the standard deviation
among different observations made at a fixed
heliocentric angle.
Fulfilling equations~(\ref{Hcond}) and (\ref{depo}) 
warrants producing the observed \linee\ Hanle 
depolarization. 
We will refer to them as the {\em Hanle constraint}.

Lacking of additional observables to  determine a single
$P(B)$, we resort to a general consideration of theoretical 
nature.
The magnetic energy associated with the quiet Sun magnetic fields
has to be smaller than the kinetic energy of the granular motions,
otherwise the non-magnetic numerical simulations of 
solar granulation
would not be so successful in reproducing all kinds
of observables
\citep[see][]{stei98,asp00}. 
In other words,
\begin{equation}
\chi\ll 1,
\label{smallchi}
\label{frac_ener}
\end{equation}
if $\chi$ is the ratio between the magnetic
energy and the kinetic energy of the granular motions
at the base of the photosphere,
\begin{equation}
\frac{\langle B^2\rangle}{8\pi}=\chi \frac{1}{2}\langle\rho u^2\rangle.
\label{kinetic}
\end{equation}
Using the typical convective velocities $u$ and densities
$\rho$ at $z=0~$km 
($\rho\simeq3\times 10^{-7}$g\,cm$^{-3}$,  $u\simeq3$~km~s$^{-1}$; 
 \citealt{stei98}, Figure~5), one gets a kinetic energy density of 
 some $1.3\times 10^{4}~{\rm erg~cm}^{-3}$. This value allows us 
 to rewrite equation~(\ref{kinetic}) as,
 \begin{equation}
 \chi=\langle B^2\rangle/(580~{\rm G})^2.
 \label{kinetic2}
 \end{equation}

%%%%%%%%%%%%%%%%%%
\section{Results}\label{results}

The PDF in equation~(\ref{eq1}) depends on three free
parameters. As it is discussed in \S~\ref{constraints},
we do not have three independent constraints to 
determine a single $P(B)$. There is one genuine
constraint (equations [\ref{Hcond}] and [\ref{depo}]) and 
two lower  limits (equations [\ref{ww}] and [\ref{frac_ener}]).
This problem rules out selecting one PDF for
the quiet Sun magnetic field strength. 
Instead,  we are forced to explore the
set  of PDFs compatible with the observables, and then 
to distill  properties common to all of them. 
The strategy of the search and the common properties
are described next.

\begin{figure*}
\resizebox{\hsize}{!}{\includegraphics{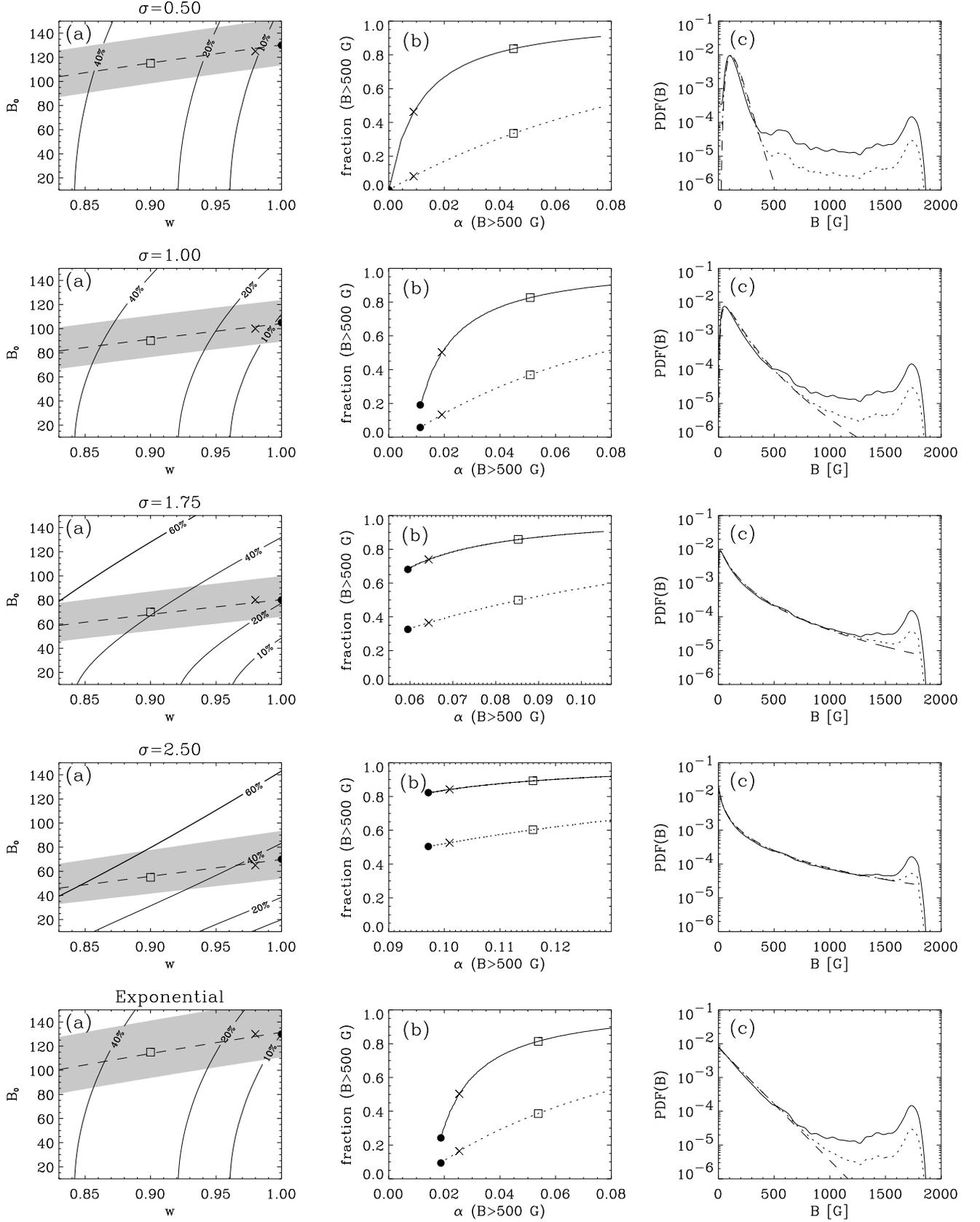}}%pdf3.ps}}
\caption{
	(a) Magnetic energy density of the PDFs as a function
	of the free parameters $w$ and $B_0$ (the solid
	contours). The labels give the magnetic energy 
	in percent of the kinetic energy of the 
	granular motions. The dashed lines show those
	pairs $w$ -- $B_0$ fulfilling the Hanle constraint, with
	the shaded strips representing to the range
	of observed depolarizations.
	The different plots portray
	different values of the parameters $\sigma$, plus
	the case of an exponential Hanle PDF included for 
	reference (bottom;
	in this case $B_0=\langle B\rangle$).
	The three symbols correspond to the three symbols
	in the second column, and to the three PDFs of the
	third column (see below). 
	(b) Fraction of quiet Sun unsigned magnetic flux density
	(the dotted line)
	and of magnetic energy density (the solid line) in the form
	of strong fields ($B> 500$~G).  They are represented
	versus their filling factor $\alpha(B > 500~{\rm G})$.
	The
	curves correspond to pairs $w$ -- $B_0$ fulfilling
	the Hanle constraint (the dashed lines of the first
	column).
	(c) PDFs corresponding to the three symbols in the first
	and the second columns. The solid line corresponds
	to the square ($w=0.9$), the dotted line to the times
	symbol ($w=0.98$), and the dashed line to the 
	bullet ($w=1.0$).
 }
\label{pdf3}
%\label{ca}
\end{figure*}

Given $\sigma$, we produce a grid of $18\,\times\, 30$
$P(B)$ with $0.83 \leq w \leq 1$, and 
$5~{\rm G} \leq B_0 \leq 150~{\rm G}$. It suffices 
to include all the possibilities allowed by the 
observational and theoretical constrains. For each PDF
we compute the total magnetic energy in units of the 
kinetic energy ($\chi$ in 
equation~[\ref{kinetic2}]), and the \linee\ Hanle signals to be expected
(equation~[\ref{Hcond}]). The magnetic energy and the
Hanle signals as a function of $w$ and $B_0$
are shown in the plots of the first column of Figure~\ref{pdf3}
(column a).
The total energy is represented by the
solid contours. The dashed lines trace those
pairs $w$ -- $B_0$
compatible with the observed depolarization~(\ref{depo}).
The shaded strips around these dashed
lines correspond to the error bars of the 
depolarization.
The five plots portray  
four  different values of $\sigma$ ($0.5, 1, 1.75,$ and 
$2.5$), plus the case of exponential Hanle
PDFs (equation  [\ref{exp}]).
The first column of
Figure~\ref{pdf3} shows that the magnetic energy 
is always important (say, larger than 15\,\% of the
kinetic energy of the granular motions). 
This conclusion turns out to be independent
of $B_0$ and $\sigma$ provided that $w$ fulfills the 
limit~(\ref{ww}).

\begin{figure}
\resizebox{\hsize}{!}{\includegraphics{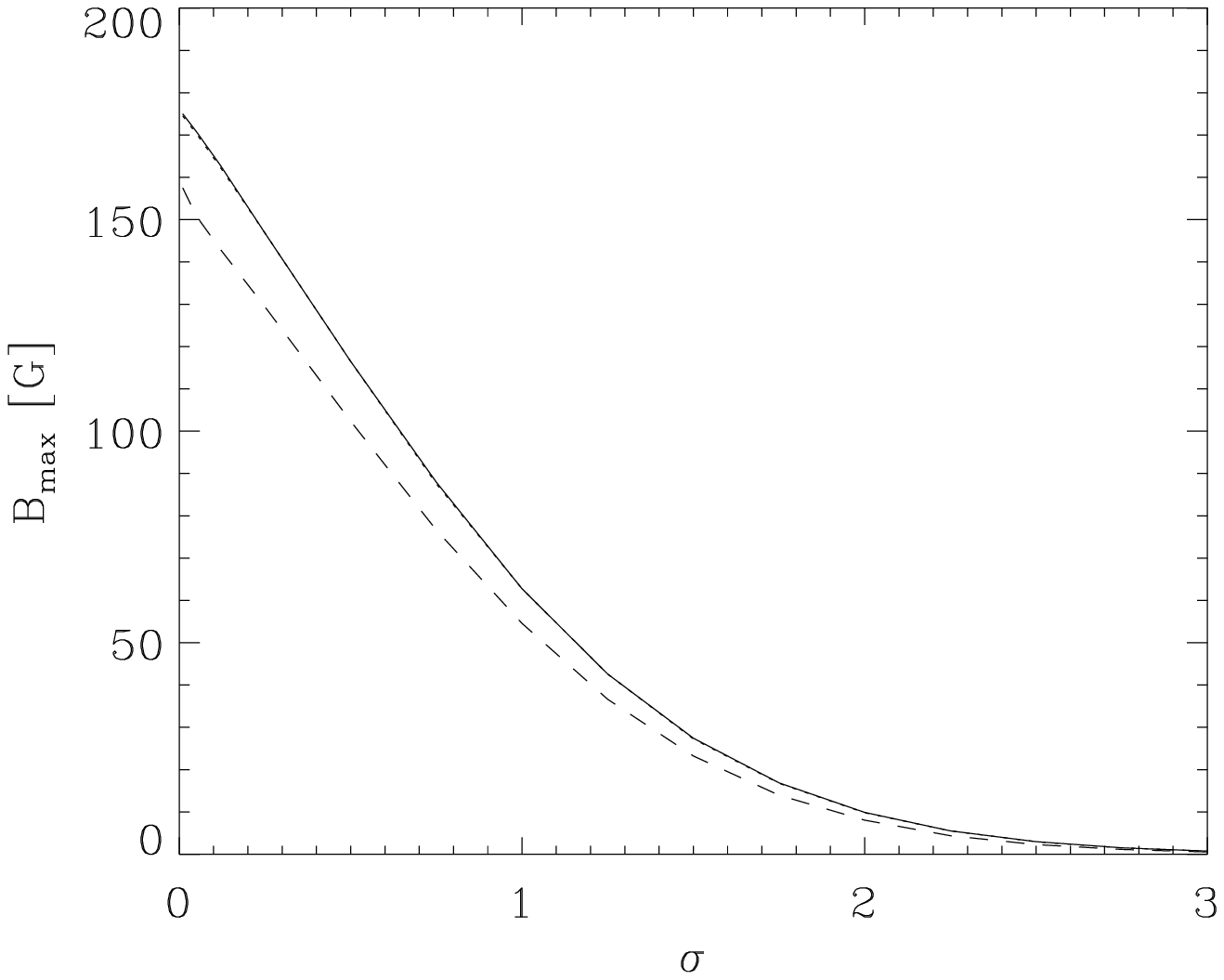}}%figmax.ps}}
\caption{
Magnetic field strength at which $P(B)$ is maximum versus
$\sigma$.
The different curves have different $w$;
1, 0.98 and 0.9, for the solid line, the (barely visible)
dotted line, and the dashed line, respectively. 
}
\label{bmaxvaries}
\end{figure}

The third column in Figure~\ref{pdf3} shows specific PDFs
for three pairs $w$ -- $B_0$ producing the observed Hanle signals.
The dashed line, the dotted line and the solid line correspond to the
circle,
the cross and the square in the first column, 
respectively. They  differ mainly because of $w$,
a parameter controlling the importance
of the tail of strong fields, with little 
influence on the core of the PDF at low field
strengths.
As we discuss in \S~\ref{shape},
the PDFs  present a 
maximum in this core of small $B$.
(The most probable magnetic field strength 
differs from zero.) 
Since the Hanle constraint determines $B_0$ 
almost independently of $w$ 
(the dashed lines in Figures~\ref{pdf3}a are nearly 
horizontal), the position of this maximum
is controlled by $\sigma$ (see equation~[\ref{maxposition}]).
Figure~\ref{bmaxvaries} shows how 
the magnetic field strength of the maximum
$B_{\rm max}$ increases
with decreasing $\sigma$ to become 
$B_{\rm max}\simeq 180$~G when $\sigma\rightarrow 0$.
This limit can  be  explained  
keeping in mind that $P(B)$ tends to
a Dirac $\delta$-function $\delta(B-B_0)$
when $\sigma\rightarrow 0$
(see equation~[\ref{logn}]).
A Dirac $\delta$-function at $B\simeq 60$~G accounts for 
the Hanle constraint~(\ref{depo}) \citep{tru04,bom05,san05}.
According
to the recipe that we use for the
variation with height of the PDFs
(equations~[\ref{var2}] and [\ref{neq3}]), a 
$\delta(B-60~{\rm G})$ 
at the height where \linee\ is formed corresponds
to $\delta(B-180~{\rm G})$ at $z=0$, which explains
the limit $B_{\rm max}\simeq 180$~G.

\begin{figure}
\resizebox{\hsize}{!}{\includegraphics{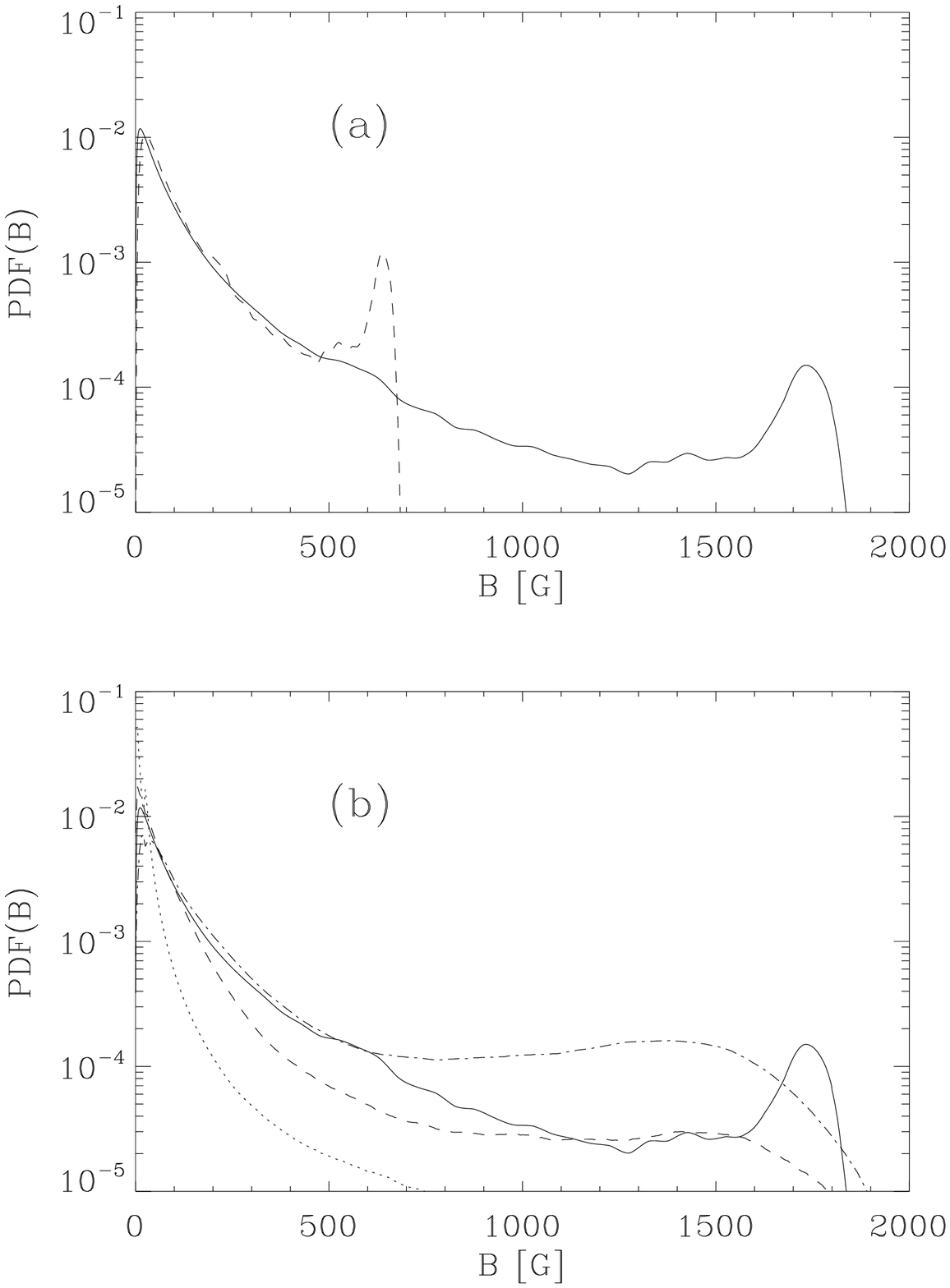}}%pdf4b.ps}}
\caption{Reference PDF with $\sigma=1.75$ and $w=0.94$. (a)
  PDF at $z=0$~km (the solid line) and  at $z=250$~km
  (the dashed line). (b) Comparison of $P(B)$ at $z=0$
  (the solid line) with PDFs from numerical simulations 
  of magneto-convection by \citet{vog03b}
  having  initial   field strengths of
  10~G (the dotted line), 50~G (the dashed line)
  and 200~G (the dotted-dashed line).
}
\label{pdf4}
\end{figure}

In order to illustrate some 
other details of the PDFs fulfilling
the observational constraints, we select a 
representative one having a few additional
desirable properties. It will be denoted 
as the {\em reference}~PDF.
In principle, there are no arguments to prefer
a specific $B_{\rm max}$, and a wide range of values
is compatible with the observables (Figure~\ref{bmaxvaries}).
However, we tend to think that small $B_{\rm max}$
PDFs are more realistic, since 
they seem to be preferred by the
numerical simulations 
of magneto-convection 
\citep{emo01,ste02,vog03b,vog05}.
Small $B_{\rm max}$ implies large $\sigma$, but
 $\sigma$ cannot be  arbitrarily large since
the magnetic energy of the PDF increases
with $\sigma$ (Figure~\ref{pdf3}a), and the magnetic
energy  must comply with the  limit~(\ref{smallchi}). 
We choose $\sigma=1.75$, which yields $B_{\rm max}\simeq 13~$G.
This selection together with  $w=0.94$ leads to the reference
PDF shown in Figure~\ref{pdf4}.
It has an energy 28\% of the kinetic energy ($\chi=0.28$), and
an unsigned flux density $\langle B\rangle\simeq$150~G.  
The dashed line in Figure~\ref{pdf4}a also shows
the reference PDF at 250~km, computed according to the
recipe in equation~(\ref{var2}). The latter has
an unsigned flux density of 120~G,
similar to the flux of the exponential PDF used by
\citet{tru04} to account for the
Hanle constraint in equation~(\ref{depo}). 
The  magnetic energy at $z=250$~km is
three times smaller than the energy
at the base of the photosphere.
Figure~\ref{pdf4}b compares our reference PDF and
simulated PDFs from \citet{vog03b} with initial fields 
of 10~G (the dotted line), 50~G (the dashed line), and
200~G (the dotted-dashed line).
These 3D simulations solve the MHD equations for 
a compressible partly ionized plasma. After a purely
hydro-dynamic transient, an initial 
homogeneous and vertical field
is introduced. The processing and amplification
of these seed fields lead to additional magnetic
flux, which amounts to 25~G, 100~G and 270~G,
for the three different initial fields. 
The reference PDF turns out to be
a mixture between the 200~G PDF ($B< 600$~G)
and the 50~G PDF ($1000~{\rm G}< B< 1500$~G). Note
that the two numerical PDFs have a small broad 
local maximum
at large magnetic fields ($B\sim$ 1500~G). Our 
semi-empirical PDF also has a maximum at large fields,
except that it is more pronounced, and it occurs 
at higher fields 
($B\simeq$ 1700~G). The existence of such maximum
at hight field strengths can be understood as the
result of a magnetic intensification
process from weak to strong fields. The 
magnetic energy piles up at the largest possible
field strength, set by the value whose magnetic
pressure exceeds the gas pressure
of a non-magnetic atmosphere in mechanical
equilibrium  ($\sim$ 1800~G at $z=0$~km).
(A detailed explanation of the
mechanism can be found in \citealt{dom05}.)

We are particularly interested in understanding which
part of the quiet Sun magnetism is provided by 
strong fields which, among other things, are responsible 
for most of the observed Zeeman signals (see Appendix~\ref{appc}),
and can be studied in unpolarized light (see below).
In order to quantify the contribution, we use the
first moments of the PDF. Specifically, the
fraction of quiet Sun occupied by strong fields, 
\begin{equation}
\alpha(B>B^*)=\int_{B^*}^\infty P(B)\,dB,
\end{equation}
the fraction of unsigned magnetic flux,
\begin{equation}
\phi(B > B^*)=
{1\over{\langle B\rangle}}\int_{B^*}^\infty B\,P(B)\,dB,
\end{equation}
and fraction of magnetic energy,
\begin{equation}
\varepsilon(B > B^*)=
{1\over{\langle B^2\rangle}}\int_{B^*}^\infty B^2\,P(B)\,dB.
\end{equation}
Choosing a somewhat arbitrary threshold between weak
and strong fields\footnote{This figure is
not far from the value of a magnetic field strength yielding
equipartition between magnetic energy density and
kinetic energy density; see 
equation~(\ref{kinetic2}) with $\chi=1$.},
we use
\begin{equation}
B^*=500~{\rm G},
\end{equation}
for 
the plots in the second column (b) of Figure~\ref{pdf3}.
They show the fraction of unsigned flux (the dotted line) and
magnetic energy (the solid line) versus the volume occupied
by the strong fields. Only PDFs producing the observed 
Hanle depolarization are considered (those corresponding to the
dashed lines in the first column of Figure~\ref{pdf3}).
The dots, crosses and squares also correspond to the
dots, crosses and squares in the first column of Figure~\ref{pdf3}. 
In all cases with $w<0.98$ (indicated by  the 
cross symbols), the fraction of
magnetic energy in strong fields is larger than 50\%
(between 45\% and 85\% when $0.5\le \sigma\le 2.5$),
although 
they fill only a small fraction of the surface 
(between 
1\% and 10\%). The contribution is 
disproportionated. As far as the unsigned
flux is concerned, the strong field still provide a sizeable fraction 
(between 10\% and 50\% for $0.5\le \sigma\le 2.5$).
The reference PDF ($\sigma=1.75$,  $w=0.94$) 
has 42\,\% of the unsigned flux in $B> 500$~G,
a figure similar to the estimates by
\citet[][38\%]{soc03}
and \citet[][36\%]{san04}.  

\begin{figure}
\resizebox{0.6\hsize}{!}{\includegraphics{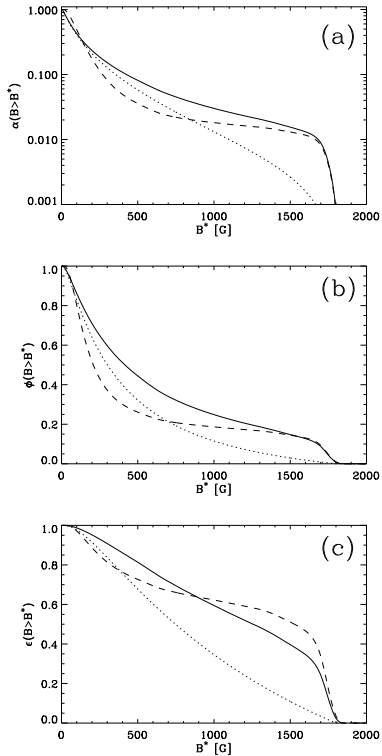}}%pdfb1.ps}}
\caption{
Filling factor (a), fraction of unsigned 
magnetic flux (b), and  fraction of magnetic energy (c) 
in field strengths stronger than $B^*$. The
three curves correspond to three PDFs fulfilling
the Hanle constraint with $\sigma=1.75$ and $w=0.94$
(the solid line,  and also the reference PDF),
 $\sigma=1.75$ and $w=1$
(the dotted line), and $\sigma=1$ and $w=0.94$
(the dashed line). 
}
\label{pdfb1}
\end{figure}

According to a physical mechanism originally
proposed by \citet{spr76}, the very strong
fields are expected to be bright in
unpolarized images.
Part of the pressure required to maintain the 
plasma in mechanical balance is provided by the magnetic
pressure. When the magnetic pressure of a
particular structure is comparable to 
the mean gas pressure of the atmosphere, the 
structure must have low mass density and so it
becomes transparent. We see deeper 
through the magnetized structure, and deeper often means
hotter and brighter. The mean pressure at $z=0$ is equivalent
to a magnetic pressure provided by a field strength
of some 1800~G.
In order to estimate which fraction of the
quiet Sun magnetic fields could be bright,
we consider the strongest magnetic fields in the PDFs,
say, those larger than two-thirds 
of the maximum value, or larger than 1200~G. 
Figure~\ref{pdfb1}a shows $\alpha(B>B^*)$
versus $B^*$ for three representative PDFs fulfilling
the Hanle constraint. 
The reference PDF corresponds to the solid line. 
The filling factor in
bright features is always very small,
e.g., $\alpha(B>1200~{\rm G})=$\,2.5\,\% in the
reference PDF. Yet, these fields are responsible for
25\,\% of the unsigned flux  and 50\,\% of the 
magnetic energy of the quiet Sun magnetic fields.
\citet{san04a} detected (G-band) bright points in the 
quiet Sun covering some 0.7\,\% of the supergranular cell
interiors.  Although the percentage is smaller than the surface
coverage in strong fields mentioned above,
the two figures may not necessarily disagree.
As it is pointed out by 
\citet{san04a}, the observed value is  strongly biased
and it represents a lower limit. Only the largest
magnetic 
concentrations pop up above the dark 
intergranule background \citep{tit96} and, 
moreover, not all kG are expected to be 
bright \citep{san01,vog03b}.

So far the value of $Q/Q_0$ given in 
equation~(\ref{depo}) has been used as the Hanle 
constraint. In order to  explore the dependence of the 
results on  $Q/Q_0$, we  also tried  $Q/Q_0\simeq 0.6$, 
which is the Hanle  depolarization inferred and
used  by \citet{fau01}. The main effect is decreasing
the value of $B_0$ needed to reproduce the observations,
and so, decreasing the magnetic field where $P(B)$ is maximum 
($B_{\rm max}$,  see equation~[\ref{maxposition}]).
This shift of $B_{\rm max}$ can be achieved
without a large variation of the
unsigned magnetic flux and magnetic
energy. This is due to the fact that the
tail of hG and kG field strengths
is the mayor contributor to the flux and
energy, and this tail is not very
sensitive to $B_0$, and so, to
$B_{\rm max}$. For instance,  the  
reference PDF  ($w=0.94$, $\sigma=1.75$) has 
$B_{\rm max}=3$~G, 
$\langle B\rangle=$70~G, and
$\chi\simeq0.17$ when $Q/Q_0\simeq 0.6$,
whereas  $B_{\rm max}=13$~G, 
$\langle B\rangle=$150~G, and
$\chi\simeq 0.28$ when $Q/Q_0\simeq 0.41$.
This moderate change in $\langle B\rangle$
and $\chi$ must be compared with the case
where the tail of strong fields
is not present
($w=1$ in equation~[\ref{eq1}]). Then 
$\langle B\rangle\propto B_{\rm max}$
and $\chi\propto B_{\rm max}^2$
(equations~[\ref{firstmoment}],
[\ref{secondmoment}],
and [\ref{maxposition}]), 
which implies a very large decrease of the
unsigned
flux (a factor of four) and energy (a factor of twenty)
to fit a depolarization signal of
$Q/Q_0\simeq 0.6$.

The unsigned magnetic flux and magnetic energy
assigned by our semi-empirical PDFs to the
quiet Sun is far larger than the unsigned 
flux and energy in the form of active 
regions and the network, even at solar maximum. 
The  reference PDF has 
$\langle B\rangle\simeq 150~{\rm G}$
and $\langle B^2\rangle^{1/2}\simeq 305$~G.
Consider the regular full disk
magnetograms, which are mostly sensitive to 
the active  regions and  the network
(see \S~\ref{intro}). At solar maximum, these
magnetograms show a total unsigned magnetic flux 
across the solar surface of some $7\times 10^{23}$~Mx
\citep[e.g.,][Chapter 12]{har93}.
Dividing this magnetic flux by the area
of the solar surface, the unsigned flux 
density in the
form of active regions and the network 
$\langle B\rangle_{AR}$ turns out to be, 
\begin{equation}
\langle B\rangle_{AR}\sim 12~{\rm G}
	\sim 0.08~\langle B\rangle,  
\end{equation}
where we have used for $\langle B\rangle$
the unsigned flux of the reference PDF.
If all these active region
magnetic fields are intrinsically
strong with $B\sim 1500$~G, then
\begin{equation}
\langle B^2\rangle_{AR}\simeq  1500~{\rm G}~\langle B\rangle_{AR}
\sim (135~{\rm G})^{2}\sim 0.2~\langle B^2\rangle .
\end{equation}

%%%%%%%%%%%%%%%%%%%%%
\section{Conclusions}\label{conclusions}

The quiet Sun magnetic fields may be central
to understand the global magnetic properties of
the Sun, but they are not well 
characterized from an empirical point of view. They 
present a variety of physical properties and so their 
description must be carried out in terms of probability 
density functions (PDFs). This work is focused on
characterizing  the PDF of the field strength, which
describes the 
fraction of quiet Sun occupied by each field 
strength.
We use the symbol $P(B)$ for the PDF at the
base of the photosphere\footnote{Where the continuum
optical depth at $\lambda$5000~\AA\ is one, and the total 
pressure $\simeq 1.2\times 10^{5}$~dyn~cm$^{-2}$.}. 
The PDFs existing in the literature provide only a partial
view  -- each  individual  
measurement samples only  a small range of 
field strengths (see \S~\ref{intro}). Our
paper proposes a semi-empirical $P(B)$ that tries 
to get rid of all the biases and so, for the
first time,  it covers all the range of
possible field strengths  from $B= 0~G$  to $B\simeq$ 1800~G.
It combines Hanle effect based measurements, 
Zeeman effect based  measurements, and some general ansatzs 
inspired by numerical simulation of magneto-convection
and the thin fluxtube approximation (\S~\ref{synthesis}). 
The procedure is of general nature and eventually,
it   will allow us to determine a single PDF characteristic
of the quiet Sun magnetic fields. So far the 
observational and theoretical constraints 
only provide  a range of PDFs. The constraints used in this work
are: (a) reproducing the Hanle depolarization
signals of \linee , (b)  having a filling factor for
strong fields larger than the filling factor 
inferred from the observed Zeeman signals, and 
(c) having a magnetic energy density  smaller than
the kinetic energy density  of the granular 
motions.

The properties common to all $P(B)$ fulfilling these
constraints are analyzed in \S~\ref{results}.  They provide
a qualitative picture of the quiet Sun magnetic
field plus some quantitative results.
\begin{itemize}
\item[-] The magnetic energy density is 
a significant fraction of the kinetic
energy density of the granular motions.
Larger than some 15\%~of the kinetic energy at the base 
of the photosphere.
\item[-] Most of the surface is occupied by
weak magnetic fields. We find that magnetic fields weaker
than 500~G fill between 90\% and 99\% 
of the surface.
\item[-] The most probable field strength is not zero.
This result has a novel consequence. 
The existence and 
position of such maximum is a direct
indication of the presence of large amounts of unsigned 
magnetic flux in the quiet Sun. 
Its presence represents a specific prediction of the
work, and it may be 
tested  observing Hanle
signals of lines with assorted sensitivities to weak 
magnetic fields.
\item[-] The magnetic energy tends to be concentrated in strong
fields independently of the parameters chosen for the PDF. 
Between 45\,\% and 85\,\% of the energy is in the form
of fields stronger then 500~G which occupy
only a small fraction of the quiet Sun
(between 1\,\% and 10\,\%) . 
The unsigned magnetic flux of these strong 
fields represent between 10\,\% and 
50\,\% of the quiet Sun unsigned flux.
Since the strong fields reach 
the upper photosphere more easily
thant the weak ones (\S~\ref{strat}),
they may be particularly influential in connection
with the chromospheric and coronal magnetism
\citep[e.g.][]{sch03b}.
\item[-] The tail of very strong kG fields of the PDFs is expected
to be bright in
high spatial resolution unpolarized
images \citep{spr76}.
\citet{san04a} detected (G-band) bright points in the 
quiet Sun supergranulation cell interiors covering 
some 0.7\,\% of the surface.
The filling factors predicted here  are
larger but not necessarily incompatible.
The observed bright points are only
a fraction of the existing ones (the largest and
brightest ones). 
Despite the minute fraction of the surface
that it covers, this tail of kG fields
contains a significant part of the quiet Sun
magnetic  energy.
They have the practical advantage with respect
to the weak fields of showing up
in unpolarized light.
\item[-] The unsigned magnetic flux and magnetic
energy assigned by our semi-empirical PDFs to the
quiet Sun are far larger than the unsigned 
flux and energy in the form of active 
regions and the network, even at solar maximum.  
\end{itemize}

The sensitivity of the current Zeeman measurements
allows us to detect spatially resolved 
uniform magnetic fields of one G
or even smaller \citep[e.g.,][]{kel94,lit96,lin99,kho02}. 
Most of the solar surface is occupied by magnetic 
fields with strengths larger than this sensitivity 
limit which, however, do not show up in the measurements
(c.f. the Zeeman inferred  PDFs
in Figure~\ref{pdfz} with the reference PDF 
in Figure~\ref{pdf4}).
The Zeeman signals seem to be  dominated by the
strong magnetic fields,  despite the fact they 
only fill a small part  of the solar surface.
As we explain in Appendix~\ref{appc}, 
the reason is twofold. First, the Zeeman signals
are sensitive to the magnetic flux, and the
unsigned magnetic flux  in the form of
weak and strong fields is comparable.
Second, all Zeeman signals tend to cancel,
but the cancellation of opposite
polarity low field strength signals
is more effective.  Typically, 
there are many more weak magnetic field features
per resolution element than strong magnetic
field features.  Then the weak fields
cancel more often than the strong fields, and the
latter are detected in a larger proportion.
(See Appendix~\ref{appc} for 
quantitative arguments.)

Our conclusions are in the vein of previous works 
assigning large amounts of unsigned magnetic flux and 
magnetic energy to the quiet Sun 
\citep{san00,dom03a,san03,san03d,san04,tru04,bom05}.
On the other hand, they seem to contradict another line of
research indicating that the quiet Sun is
not so magnetically active 
\citep{fau93,wan95,fau01,kho05}.
As we discuss in \S~\ref{intro}, all previous estimates
are known to be biased, and so, they require 
model-dependent corrections to go from the 
actual measurements to the solar magnetic fluxes
and energies. Possibly a significant part of the 
discrepancy can be pinned down to such 
corrections, and it will be cured upon 
refinement of the techniques and improvement
of the statistical significance of the results.  

A quiet Sun as magnetic as the one advocated
here may pose a problem of consistency
with existing hydro-dynamic
numerical simulations of convection 
\citep[e.g.,][]{spr90}. The simulations
reproduce with realism a large number of observables
\citep[e.g.,][]{stei98,asp00}, even though
they do not include magnetic fields.
Could the quiet Sun have large 
amounts of magnetic flux and energy
without modifying the convective pattern
in a significant way?
In other words, what are the minimum unsigned flux and 
energy required to alter the convective pattern?  
These are open questions
to be addressed with realistic
numerical simulations
of magneto-convection.

On final comment is order. 
We simplify the problem
of characterizing the quiet Sun magnetic field strengths
using a single representative PDF. However,  it is to be 
expected that different constituents
of the quiet Sun have very different PDFs. For example,
we do not distinguish
between the PDFs of  
granules and intergranules although both numerical 
simulations \citep{cat99a,vog03b} and measurements 
\citep{soc04,tru04} indicate a clear preference
of the strong fields for the intergranules. 
The separation
between  the PDFs in granules and intergranules
remains pending of future work. 
Similarly, the Zeeman PDF adopted 
in \S\ref{shape} is taken from 
a specific observation, and it
may depend on the spatial 
resolution and other specificities 
of the particular observation.
Studying such dependence remains pending too.

\acknowledgments

Thanks are due to Alexander V\"ogler for providing the 
numerical PDFs used in the work and
shown in Figures~\ref{pdf1} and \ref{pdf4}.
Thanks are also due to the referee for helpful comments.
The work has been partly funded by the Spanish Ministry of Science
and Technology, project AYA2004-05792, as well as by  Swiss
International
Space Science Institute.

%%%%%%%%%%%%%%%%%%%%%%%%%%%%%%%%%%%%%%%%%%%
%%             APPENDIXES                %%
%%%%%%%%%%%%%%%%%%%%%%%%%%%%%%%%%%%%%%%%%%%

\appendix
\section{The unbiased  PDF is a linear combination of 
biased PDFs coming from Zeeman and Hanle measurements}
\label{ap1}

%Full derivation.
The spectra used to derive magnetic field
strengths from the Zeeman polarization signals
have a finite spatial resolution. Therefore,
two polarities often
co-exist in the resolution elements, which cancels
part of  the polarization signals. As a
result, the Zeeman  measurements 
provide only a fraction $f(B)$ of the
quiet Sun magnetic fields with strength $B$. The complementary $1-f(B)$
	is interpreted as field free. Using the symbols
	$P(B)$ for the true PDF, $\tilde P_Z(B)$ for the 
biased
PDF inferred from Zeeman signals, and $\delta(B)$ for
a Dirac $\delta$-function,
\begin{equation}
\tilde P_Z(B)=P(B)\, f(B)+
[1-\langle f\rangle ]\,\delta(B),
   \label{zeeman}
\end{equation}
with $\langle f\rangle$ the fraction of magnetized quiet
Sun according to the biased Zeeman measurements,
\begin{equation}
\langle f\rangle =\int_0^{\infty} f(B)\,P(B)\, dB,
\label{defmeanf}
\end{equation}
and $\tilde P_Z(B)$ normalized to unity.
	We ignore the details of how the fraction $f(B)$
	depends on the field strength $B$. However, it is to be
	expected that the bias of the Zeeman
	PDF increases toward low field strengths
	because the
	polarization signals weaken with decreasing field strength
	\citep[e.g.][]{unn56,lan73}.
	In addition, the bias $f(B)$ should not vary
	with the field strength
	for magnetic fields stronger than a certain limit $B_1$.
	We conjecture that given a spatial resolution, the cancellation
	of polarization signals  becomes independent
	$B$  when $B> B_1$. One
	can summarize these general conditions to be
	satisfied by $f(B)$ as,
	\begin{equation}
	f(B)\simeq\cases{
		\ll f_1&$B\rightarrow 0$,\cr
		f_1& $B> B_1$.
		}
	\label{cond1}
	\end{equation}
	Likewise, the PDF derived from Hanle signals, $P_H(B)$,
	also differs from the 
	true PDF. The difference is parametrized by $g(B)$,
	\begin{equation}
	   P_H(B)={{g(B)}\over{\langle g\rangle }}P(B).
	   \label{hanle}
	\end{equation}
	The function $g(B)$ is also unknown, except that it has 
	to allow $P_H(B)$ to be a good approximation of $P(B)$
	at small field strengths. In addition, the Hanle signals
	are insensitive to field strengths larger than the
	so-called  Hanle saturation $B_2$ (see \S~\ref{intro}).
	This fact allows us to assume $g(B)\simeq 0$ for
	$B> B_2$  since no information
	on the true PDF  is provided by $P_H(B)$,
	\begin{equation}
	g(B)\simeq \cases{1& $B\rightarrow 0$,\cr
		 0 & $B> B_2$.
		} \label{cond2}
	\end{equation}
	For the sake of convenience, we consider  
	$\tilde P_Z(B)$ without the $\delta$-function. After
	normalization, the process renders a new Zeeman PDF,
	\begin{equation}
	P_Z(B)=\big[\tilde P_Z(B)-
	\big(1-\langle f\rangle \big)\,\delta(B)\big]\,\langle f\rangle ^{-1}
	={{f(B)}\over{\langle f\rangle }}P(B).
	\end{equation}
	In principle, one could infer
	the value of $P(B)$ from both the
	Zeeman PDF,
	\begin{equation}
	P(B)=P_Z(B)    {{\langle f\rangle }\over{f(B)}},
		\label{this}
	\end{equation}
	and the Hanle PDF,
	\begin{equation}
	P(B)=P_H(B)    {{\langle g\rangle }\over{g(B)}}.
		\label{that}
	\end{equation}
	In fact, any linear combination of equations~(\ref{this})
	and (\ref{that}) also renders the true PDF,
	\begin{equation}
	P(B)=h(B)\, P_Z(B)    {{\langle f\rangle }\over{f(B)}}+
	[1-h(B)]\,P_H(B)    {{\langle g\rangle }\over{g(B)}},
	\end{equation}
	with a free weighting function $h(B)$.
	If one chooses $h(B)$ to be,
	\begin{equation}
	h(B)={{f(B)}\over{f_1}},
	\end{equation}
	then,
\begin{equation}
	P(B)={{\langle f\rangle }\over{f_1}}~P_Z(B) +
	[1-{{f(B)}\over{f_1}}]{{\langle g\rangle }\over{g(B)}}~P_H(B),
\end{equation}
or adding and subtracting $\langle g\rangle P_H(B)$,
\begin{equation}
	P(B)={{\langle f\rangle }\over{f_1}}P_Z(B) +
	\langle g\rangle P_H(B)+
	\big[1-{{f(B)}\over{f_1}}-g(B)\big] 
	P(B).
	\label{eqa12}
\end{equation}
The third term of this general expression is
negligible for both  small
and large $B$, i.e.,
\begin{equation}
1-f(B)/f_1-g(B)<< 1.
\end{equation}
When $B$ is large then $g(B)\simeq 0$ and $f(B)\simeq f_1$ (see
equations~[\ref{cond1}] and [\ref{cond2}]).
The same happens for small $B$  since $f(B)/f_1<< 1$
and $g(B)\simeq 1$ 
(equations~[\ref{cond1}] and [\ref{cond2}]).
Continuity arguments suggest that the term
also vanishes for intermediate strengths, so that
\begin{equation}
	P(B)\simeq {{\langle f\rangle }\over{f_1}}P_Z(B) +
	\langle g\rangle P_H(B).
	\label{eqa14}
\end{equation}
Using the normalization condition,
\begin{equation}
\int_0^\infty P(B)\,dB=\int_0^\infty P_Z(B)\,dB=
	\int_0^\infty P_H(B)\,dB=1,
\end{equation}
the  expression~(\ref{eqa14}) can be rewritten as,
\begin{equation}
P(B)\simeq w\,P_H(B)+(1-w)\,P_Z(B),
\end{equation}
with $w=1-\langle f\rangle /f_1$. 
Note that,
\begin{equation}
w < 1-\langle f\rangle ,
 \label{comment_eq}
\end{equation}
since $f_1 < 1$. Then the weight $w$ has to be
smaller than one minus the filling factor that
the Zeeman signals assign to the magnetic fields
of the quiet Sun 
(equation~[\ref{defmeanf}]). This constraint is used in the
main text.

In short, with very unspecific assumptions on the
biases affecting the observed  Zeeman and Hanle PDFs,
one can approximate the true PDF
as a linear combination of the observed  biased Zeeman
 and Hanle PDFs.

%%%%%%%%%%%%%%%%%%%%%%%%%%%%%%%%%%%%%%%%%%%%%%%%%%%%%%%%

\section{Why ubiquitous weak fields do not produce
 significant Zeeman signals.}\label{appc}

The sensitivity of the current Zeeman measurements
would allow us to detect a magnetic field 
of one G or even smaller if it is uniform and 
fills completely the resolution element
\citep[e.g.,][]{kel94,lit96,lin99,kho02}. 
According to the arguments in the main text, most of the
solar surface is occupied by magnetic fields with strengths
larger than this sensitivity limit which, however,
do not show up in the 
Zeeman
measurements (c.f. the measured
Zeeman PDFs in Figure~\ref{pdfz} with the 
unbiased PDF in Figure~\ref{pdf4}).
Actually,
the Zeeman signals  seem to be  dominated by the polarization
created by strong magnetic fields, despite the fact 
they only fill a small fraction of the solar surface.
Signals from opposite polarities  existing in the
resolution element cancel out, an effect 
invoked to explain the difference between the
real and the observed PDF
(Appendix~\ref{ap1}). Why does this bias
affect preferentially the weak fields?
Why do the small magnetic field strength
signals cancel  more efficiently than those 
from large field strengths?

\begin{figure}
\resizebox{0.5\hsize}{!}{\includegraphics{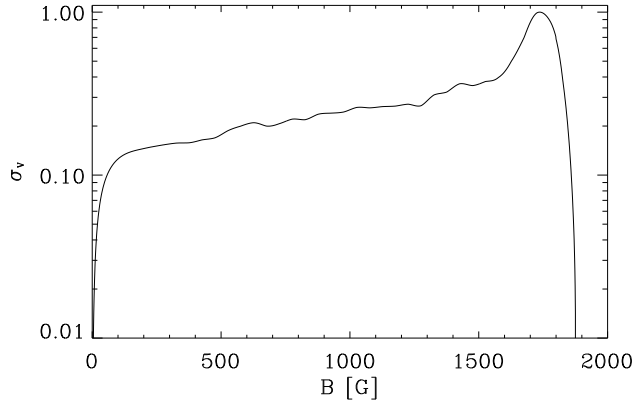}}%pdf5.ps}}
%\plotone{pdf5.ps}
\caption{Typical Zeeman signals to be 
expected for each field strength $B$. Despite
the fact that the weak fields overwhelmingly 
dominate the distribution of field strengths, they leave 
very small signals. This bias is partly due to the scaling of 
the Stokes~$V$  with the field strength, but it is also
due to a preferential canceling of the mixed polarity 
weak fields. 
The ordinates have been scaled so that the maximum equals 
one.}
\label{expected_signals}
\end{figure}

The observed bias  can be understood if the
size of the unresolved magnetic structures
contributing to the Zeeman signals is uncorrelated 
with  the magnetic field strength. (The typical
size of a unipolar structure does not depend very
much on its field strength.) Then, typically, each
resolution element contains far more structures 
having  weak
fields than structures with strong fields.  
The cancellation is more effective when there are
many structures that cancel signals, which
explains the bias.
This qualitative argument can be quantified as follows.
A unipolar patch produces a circular polarization
Stokes~$V$ signal  which,
to first order, scales with the field 
strength\footnote{The linear polarization signals
scale with the square of the magnetic field strength
and, therefore,  the bias toward large field strengths
discussed in this appendix is even more severe
in linear polarization.}.
Then the  Stokes $V$ signal to be expected in a finite 
resolution element due to magnetic fields of strength 
$B$ is
\begin{equation}
V=V_0\, B\, \mu(B)\, \big[N^+(B)-N^-(B)\big],
\label{v0}
\end{equation}
with $N^+(B)$ and $N^-(B)$ the number of structures 
with positive and negative polarities
in the resolution element.
The symbol $V_0$ stands for an irrelevant constant,
whereas $\mu(B)$ represents the typical value of the 
cosine of the magnetic field inclination with respect
to the vertical direction. (We consider observations
made at the disk center.)
The number $N^+(B)$ can be regarded as a random variable,
in the sense that it has  different values
in different resolution elements. 
Since it is a number of counts, the random variable
$N^+(B)$ follows a Poisson distribution \citep[e.g.,][]{mar71},
whose expected value and variance are identical,
\begin{equation}
{\sc E}\{N^+(B)\}=\vee\{N^+(B)\}.
\label{eevv}
\end{equation} 
Moreover,
\begin{equation}
P(B)={\sc E}\{N^+(B)+N^-(B)\}(l/L)^2,
	\label{pepe}
\end{equation} 
with $L$ and $l$ the size of the resolution element and
size of one of the magnetic elements, respectively.
(The factor 
$[L/l]^2$ is the number
of structures required to cover a resolution element.)
There is no preferred polarity, consequently, 
\begin{equation}
{\sc E}\{N^-(B)\}={\sc E}\{N^+(B)\}.
\label{pepe2}
\end{equation} 
Equations~(\ref{pepe}) and (\ref{pepe2}) lead to 
\begin{equation}
E\{N^+(B)\}={{P(B)}\over{2}}(L/l)^2.
\label{expected}
\end{equation}
If the random variables $N^+(B)$ and $N^-(B)$ are independent,
one can readily
compute the mean signal and the variance to be expected
due to the imperfect cancellation of the two
polarities. Using equations~(\ref{v0}), (\ref{eevv}),
(\ref{pepe2}), and
(\ref{expected}),
\begin{equation}
E\{V\}=V_0\, B\, \mu(B)\, \big[E\{N^+(B)\}-E\{N^-(B)\}\big]=0,
\end{equation}
whereas,
\begin{equation}
\vee\{V\}=[V_0\, B\, \mu(B)\,]^2\, (\vee\{N^+(B)\}+\vee\{N^-(B)\})=
[V_0\, B\, \mu(B)\,L/l]^2\, P(B).
\end{equation}
The typical Stokes $V$ signals to be expected would
be of the order of standard deviation of the
distribution of signals $\sigma_V$,
\begin{equation}
\sigma_V\equiv\sqrt{\vee\{V\}}=(V_0L/l)\, B\, \mu(B)\sqrt{P(B)}.
\end{equation}
Note that the Zeeman polarization signals due to $B$ do not scale
with their filling factor $P(B)$ but with $\sqrt{P(B)}$. This
scaling favors the contribution of the improbable magnetic 
fields, that is to say, the contribution of the large
field strengths. 
Figure~\ref{expected_signals}
shows $\sigma_V$ normalized so that the 
maximum equals one. It uses the reference
$P(B)$ in Figure~\ref{pdf4} plus the ansatz,
\begin{equation}
\mu(B)={{1}\over{4}}\tanh\big[(B-750~{\rm G})/300~{\rm G}\big]+
	{{3}\over{4}}
	\simeq\cases{1/2& $B <$ 500~G,\cr
	1&$B > $1000~G,}
	\label{ansatz}
\end{equation}
although the latter is secondary since
$\mu(B)$ only varies from 1/2, for an isotropic
distribution of magnetic field, to 1, for 
vertical fields. Equation~(\ref{ansatz}) parameterizes
our theoretical prejudices; the distribution of inclinations
should be more isotropic for weak fields (which can be
easily tilted by the drag of the granulation flows) and
more vertical for the strong fields (for which the
buoyancy is important; see, e.g., \citealt{sch86}).  
Figure~\ref{expected_signals} shows how 
the magnetic fields occupying most of the surface (say,
$B < 200$~G; Figure~\ref{pdf4}) leave Stokes~$V$ signals smaller than
those of the kG fields. 

In short, it is not difficult to explain why the 
Zeeman signals tend to detect strong fields, despite 
the fact that they occupy a small fraction of the solar
surface. It is due to the scaling of the Stokes~$V$ signals 
with  the field strength, plus the random cancellation 
of the mixed polarity. The latter preferentially affects the
most probable field strengths, i.e., the smallest ones. 

%%%%%%%%%%%%%%%%%%%%%%%%%%
%
% Extra info 
%\begin{figure}
%\resizebox{0.4\hsize}{!}{\includegraphics{mysurvival.ps}}
%\caption{Prepared by Ita on request. The simulations 
%have $B=50$~G, and $H_B=350$.
%}
%\label{mysurvival}
%\end{figure}

%%%%%%%%%%%%%%%%%%%%%%%%%%

%\bibliography{/home/jos/texto/papers/sun}
%\bibliographystyle{apj}

%%%%%%%%%%%%%%%
%\input{ms.bbl}

\end{document}